\documentclass[pdflatex,sn-mathphys]{sn-jnl}

\theoremstyle{thmstyleone}%
\usepackage{caption}
\usepackage{subcaption}
\theoremstyle{thmstyletwo}%

\theoremstyle{thmstylethree}%
\newcommand{\CSIVA}{CSIvA}

\begin{document}

\title[Learning to Discover Gene Regulatory Networks]{DiscoGen: Learning to Discover Gene Regulatory Networks}


\author*[1]{\fnm{Nan Rosemary} \sur{Ke}}\email{nke@google.com}

\author[1]{\fnm{Sara-Jane} \sur{Dunn}}

\author[1]{\fnm{Jorg} \sur{Bornschein}}

\author[1]{\fnm{Silvia} \sur{Chiappa}}

\author[1]{\fnm{Melanie} \sur{Rey}}

\author[1]{\fnm{Jean-Baptiste} \sur{Lespiau}}

\author[1]{\fnm{Albin} \sur{Cassirer}}

\author[1]{\fnm{Jane} \sur{Wang}}

\author[1]{\fnm{Theophane} \sur{Weber}}

\author[1]{\fnm{David} \sur{Barrett}}


\author[1]{\fnm{Matthew} \sur{Botvinick}}

\author[1]{\fnm{Anirudh} \sur{Goyal}}

\author[2]{\fnm{Mike} \sur{Mozer}}

\author[1]{\fnm{Danilo} \sur{Rezende}}

\affil[1]{\orgdiv{DeepMind}}
\affil[2]{\orgdiv{Google Brain}}




\abstract{
Accurately inferring Gene Regulatory Networks (GRNs) is a critical and challenging task in biology. GRNs model the activatory and inhibitory interactions between genes and are inherently causal in nature. To accurately identify GRNs, perturbational data is required. However, most GRN discovery methods only operate on observational data. Recent advances in neural network-based causal discovery methods have significantly improved causal discovery, including handling interventional data, improvements in performance and scalability. However, applying state-of-the-art (SOTA) causal discovery methods in biology poses challenges, such as noisy data and a large number of samples. Thus, adapting the causal discovery methods is necessary to handle these challenges. In this paper, we introduce DiscoGen, a neural network-based GRN discovery method that can denoise gene expression measurements and handle interventional data. We demonstrate that our model outperforms SOTA neural network-based causal discovery methods.
}




\maketitle

\section{Introduction}\label{intro}

Accurate inference of Gene Regulatory Networks (GRNs) is a long-standing and important problem in biology, given the potential to expose and understand the regulatory circuitry that drives cellular decision-making. These biological networks capture the regulatory relationships between transcription factors and their target genes, and aim to recapitulate gene expression dynamics. Indeed, many examples have explored putative GRNs to explain cell fate identity over differentiation trajectories, such as \cite{hartmann2019,davidson2002,dunn2014,krumsiek2011,kamimoto2023}. Beyond this, they are a powerful tool to perform in silico tests that guide experimental exploration, particularly for the goal of engineering cell fate through targeted genetic perturbation \cite{kamimoto2023dissecting}. 

It is common practice to measure gene expression at single cell resolution via single cell RNA-sequencing (scRNA-seq) \cite{zappia2021}. High-throughput gene perturbations can also be captured and measured in single cells through techniques such as Perturb-seq \cite{dixit2016}. While cost is a limiting factor that dictates the number of samples, scRNA-seq techniques can yield measurements for hundreds of thousands of cells and tens of thousands of genes. 
Such observational and perturbational datasets represent a rich source of information of cellular heterogeneity that could be exploited to learn regulatory relationships. A drawback, however, is that the datasets are noisy and sparse due to technical noise from gene dropouts that are typically brought about by low capture efficiency or amplification bias \cite{hicks2017}. This introduces a number of artificial zero measurements. Furthermore, batch effects are introduced as groups of cells get captured and measured together, introducing gene expression signatures that artificially separate similar cells. Methods that seek to learn from the richness of single cell data must also navigate such a noise that obfuscates true biological signal. 

While the value of GRN inference is rarely disputed, the tools to perform network inference demonstrate variable performance, 
and are limited in scale and compute requirements \cite{pratapa2020benchmarking}. In particular, methods that scale to large datasets are rare \cite{gibbs2022, van2020}, and nearly all are limited in being able to infer GRNs from perturbational data. This renders the GRN challenge as relevant as it is ever been. 

At its core, GRN inference requires discovering causal structure between genes \cite{chan2017gene,qiu2020inferring,belyaeva2021dci}. Recent years have seen considerable advances in algorithms for causal discovery, which describe causal structure in the form of a directed graph \cite{ke2019learning,brouillard2020differentiable,lippe2021efficient,scherrer2021learning,ke2022learning}. 
Despite progress, most algorithms assume acyclicity of the graph, are limited in the number of nodes that can be considered, and only use observational data. 
One notable exception is \CSIVA~\cite{ke2022learning}, a supervised neural-network-based method that has demonstrated state-of-the-art performance on various datasets and tasks. The ability to deal with large and cyclic graphs, and to use interventional data, renders \CSIVA~well-suited to GRNs. However, the high level of technical noise makes it challenging to use this method with scRNA data \cite{dibaeinia2020sergio}.

\begin{figure}[t]
    \centering
    \includegraphics[width=0.99\textwidth]{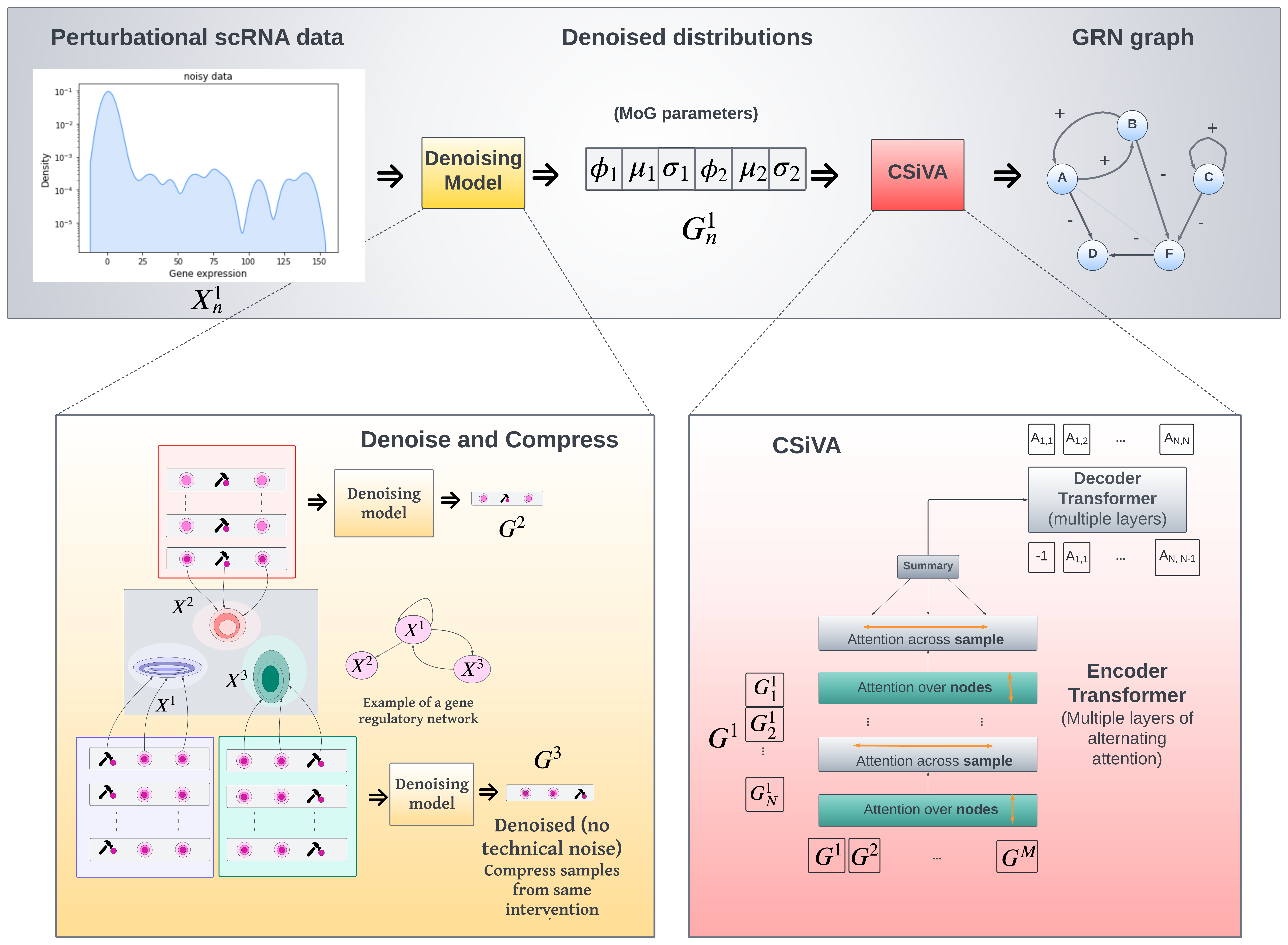} 
    \caption{Overview of DiscoGen. A denoising model takes perturbational scRNA data and compresses it into estimated technical noise-free distributions modelled as Mixture of Gaussians (MoG). The MoG parameters are then passed to \CSIVA, which outputs the GRN graph as an adjacency matrix. 
    }
    \label{fig:DiscoGen}
\end{figure}

In this work we propose DiscoGen: a framework to infer GRNs from perturbational scRNA-seq data, which combines \CSIVA~with a denoising and compression model (Figure \ref{fig:DiscoGen}). DiscoGen returns a directed graph with nodes representing genes and edges representing causal influences, which are either excitatory or inhibitory. Given noisy scRNA-seq data, the denoise and compress model is used to provide an estimate of the technical noise-free distribution for all perturbed genes. These estimated distributions are provided as input to \CSIVA, which outputs the associated GRN graph. Both models are trained in a supervised way using synthetically generated data and can be tested on any observational or perturbational single cell data. 
We show that DiscoGen significantly outperforms state-of-the-art methods \cite{ke2019learning, brouillard2020differentiable,lopez2022large} for GRN discovery on simulated GRN data. 

\section{Model Architecture} 

\paragraph*{Denoise and Compress Method}
In contrast to current gene expression denoising approaches that aim to obtain technical noise-free data \cite{eraslan2019single,patruno2021review,zhang2022ideas}, our goal is to estimate technical noise-free distributions, as this enables us to input compressed information to the causal discovery method, which greatly reduces computational burden. 

We model the technical noise-free interventional distribution associated to the perturbation of gene $X_n$ as a product of Mixtures of Gaussians (MoG) over the remaining genes of the GRN $\{X_1,\ldots,X_N\}\backslash X_n$. We estimate each MoG using a long-short-term-memory (LSTM) \cite{hochreiter1997long} and a mixture density network (MDN) \cite{bishop1994mixture}; the first is a recurrent neural network capable of learning sequence prediction problems, while the second is a neural network for modelling conditional density functions. The LSTM module takes as input scRNA data corresponding to the expression levels of genes $\{X_1,\ldots,X_N\}\backslash X_n$ resulting from $K_n$ perturbations of gene $X_n$. 
The LSTM's last hidden state is then passed to a MDN module that outputs the 6 parameters (means, variances, and mixture weights) of a MoG with 2 mixtures. The model is trained on artificial data generated with the Sergio simulator as detailed in Section \ref{sec:data}. 
See Figure \ref{fig:denoised} for a visualization of the noisy, denoised and clean data. Data drawn from a randomly selected test sample. As shown in the figure, the denoised data (sampled from the denoised data distribution) is a lot closer to the clean data distribution. Both the denoisd and the clean data distribution are significantly different from the noisy distribution. This showcase the strong performance of our denoised model.

\paragraph*{\CSIVA-based Causal Discovery Method}
Our causal discovery method is a modification of the \CSIVA~method introduced in \cite{ke2022learning}. 
\CSIVA~is a supervised method based on a transformer neural network \cite{vaswani2017attention,devlin2018bert}, a class of neural networks that use self-attention to weight the importance of different parts of the input sequence. Specifically, \CSIVA~uses a modification of a vanilla transformer that alternates attention \cite{bahdanau2014neural} between the input dimensions.

The transformer neural network takes as input a matrix of size $N\times 6M$, where $N$ is the number of genes in the GRN, and $6M$ is the number of interventional distributions parameters. More specifically, each row in the matrix corresponds to the 6 parameters of $M$ MoG distributions, each corresponding to perturbing a gene. The network outputs the GRN graph as an adjacency matrix with values 0, 1, and -1 indicating absence of an edge, presence of an excitatory edge, and presence of an inhibitory edge, respectively. As for the denoising model, the causal discovery model is trained on artificial data generated with the Sergio simulator.

\section{Data}
\label{sec:data}
To train and evaluate DiscoGen, we used in silico data generated using the Sergio GRN simulator \cite{dibaeinia2020sergio}. 
Sergio models gene expression data over time using stochastic differential equations (SDE).  
Unlike an alternative simulator, BoolODE \cite{pratapa2020benchmarking}, gene expression values are modelled as continuous, rather than binary, using a sigmoidal non-linearity, resulting in more realistic data. Furthermore, Sergio includes three types of technical noise (dropout, library size, and outlier) for 14 pre-specified levels curated according to real experimental data. We used the DS6 noise setting, as this matched against the human PBMC (10X chromium) cells \cite{dibaeinia2020sergio}. 
As Sergio does not provide graph structures, we generated random scale-free (SF) graphs of different sparsity levels, as scale-free networks are believed to exhibit topological properties similar to gene networks \citep{barabasi1999emergence}. The networks each have $N=100$ nodes, for both training and evaluation. We considered SF graphs with an average degree of 1, 2 and 3 per gene. 
We then generated steady-state perturbational datasets by performing in silico knock-out perturbations on all genes in the graph, from $K_n=100$ to $K_n=500$ replicates (i.e. cells/samples) for each perturbation\footnote{We also performed experiments for different numbers of replicates per perturbation, and found that perturbations on $100$ cells doesn't degrade performance much.}. We trained DiscoGen on $50,000$ synthetic datasets generated in this way, and evaluated using $128$ datasets. Details are given in Section \ref{sec:exp_results}.

\section{Experimental results}
\label{sec:exp_results}

We compared the performance of DiscoGen with the DCDI method \cite{brouillard2020differentiable,lopez2022large}. As evaluation metrics we used area under precision-recall curve (AUPRC), area under precision-recall curve ratio (AUPRC ratio),  percentage of edge inaccuracy  and F1 score as commonly done in the literature \cite{papili2018sincerities,pratapa2020benchmarking,dibaeinia2020sergio,lopez2022large}. 

\textbf{Baseline comparisons.} Most GRN inference methods, such as Genie3 \cite{huynh2010inferring}, PIDC \cite{chan2017gene}, Sincerities \cite{papili2018sincerities}, and SCODE \cite{matsumoto2017scode}, are not capable of handling perturbational data, which is required to discover causal structure \cite{eberhardt2012number}. 
The only baseline method that can use perturbational data as well as a large number of genes is DCDI \cite{brouillard2020differentiable,lopez2022large}, a neural-network-based causal discovery method that builds upon the Structure Discovery from the Interventions \cite{ke2019learning} and DAGs-with-NO-TEARS \cite{zheng2018dags} methods. 
However, unlike DiscoGen, DCDI assume acyclicity, which is violated in GRNs.  DCDI also requires separate training for each dataset, making it inefficient.
Hence, we evaluated DCDI on a subset of 24 randomly selected datasets from the 128 used to evaluate DiscoGen.

To facilitate a fair comparison, we conducted extensive hyperparameter sweeps for DCDI, extending beyond the ranges considered in the original paper \cite{brouillard2020differentiable}. We report the best results found using the hyperparameter sweep. Further details can be found in Table \ref{tab:app:hyper_dcdi} of Appendix \ref{sec:app:Hyper}.

For DiscoGen evaluated the methods on both the ability to infer edge direction and activation/inhibition.  For DiscoGen, we considered an edge to be correctly identified only if both its direction and type were correctly predicted. DCDI is not capable of predicting activation/ inhibition. Hence, for DCDI, we only evaluate the model's ability to infer edge directions, which is an easier task compared to the task for DiscoGen.

\begin{figure}
    \centering
    \includegraphics[width=0.6\textwidth]{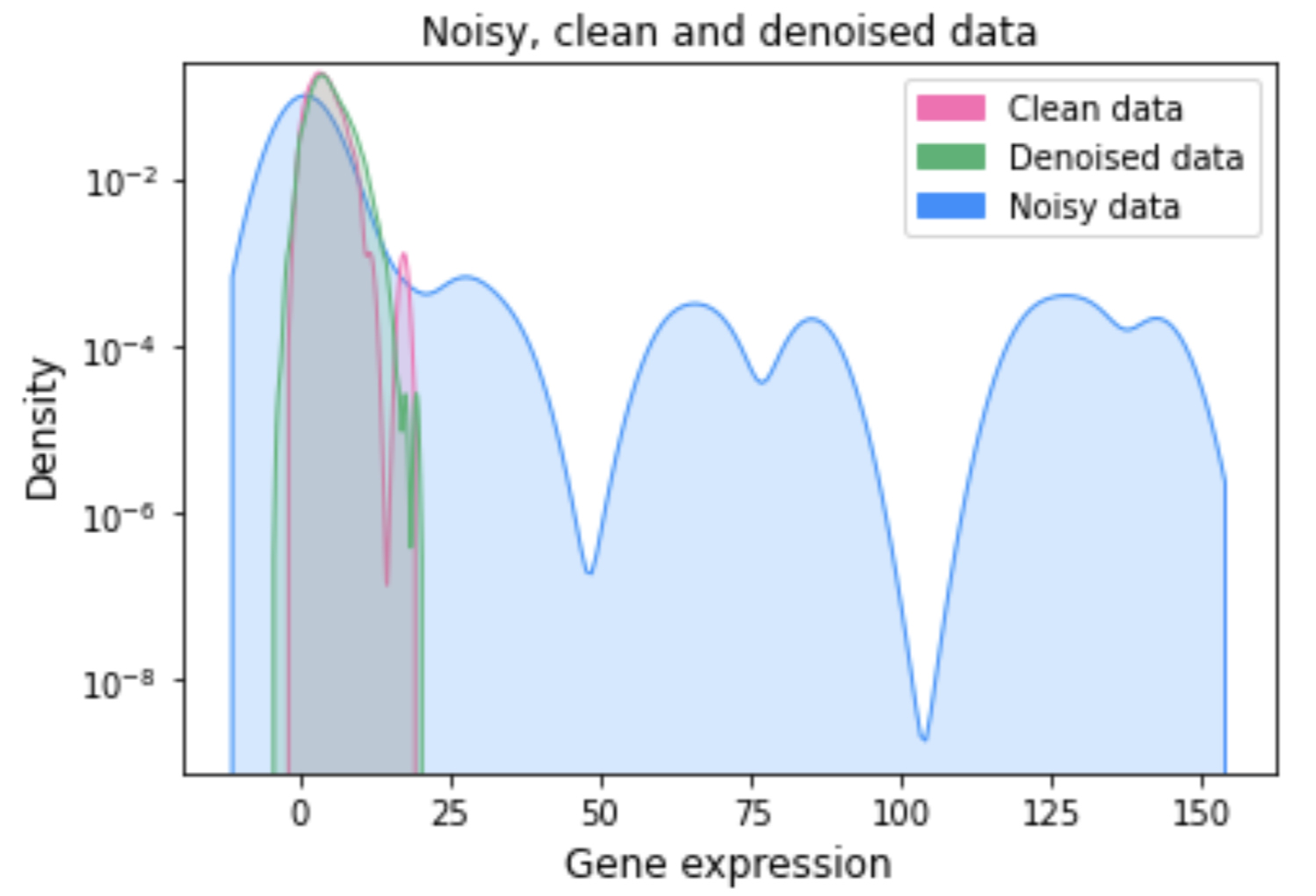} 
    \caption{Kernel density plot of clean, denoised and noisy data. Data drawn from a randomly selected test sample. The denoised data are $500$ samples from the denoised data distribution. Noisy data distribution is significantly different from the clean and the denoised data distribution. The denoised data is very close to the clean data.  }
    \label{fig:denoised}
\end{figure}

\subsection{Denoising Results}
First, to understand the performance of the denoise and compress method, we visualize data before and after denoising. Figure \ref{fig:denoised} shows data samples from a single perturbation (with $K_n=500$ replicates) without technical noise (left) alongside samples from the same perturbation after adding the DS6 technical noise using Sergio (right). We then present samples from the estimated MoG distribution. All results are randomly selected from the test set. We can see that the denoised samples are very close to the clean data, and both are markedly different from the noisy data. 

\subsection{GRN Discovery Results}

\begin{figure*}[t!]
\centering
\begin{subfigure}[t]{0.4\textwidth}
\vskip 0pt
       \includegraphics[width=\textwidth]{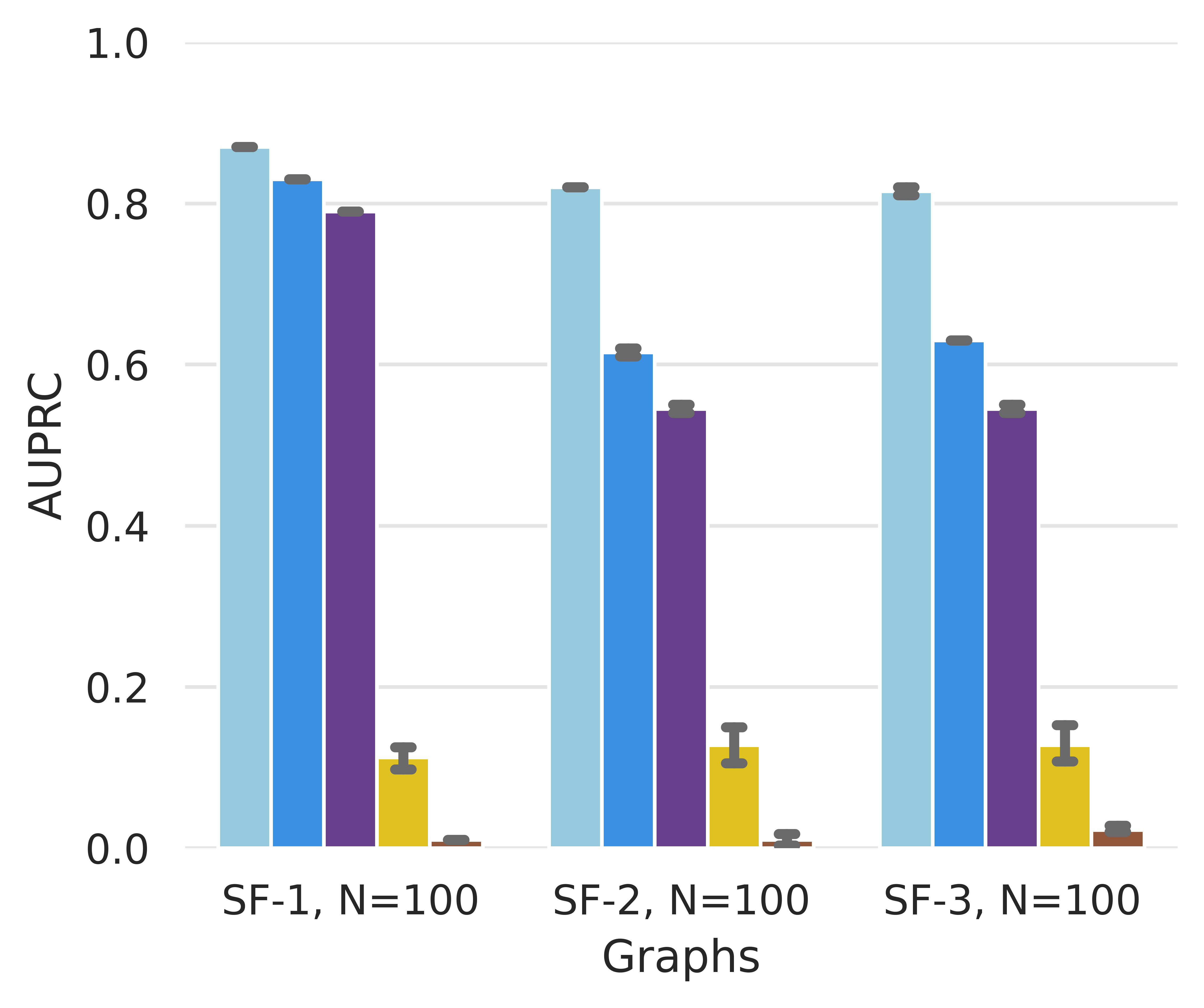}
        \caption{}
    \end{subfigure}
\hfill%
\begin{subfigure}[t]{0.4\textwidth}
\vskip 0pt
        \includegraphics[width=\textwidth]{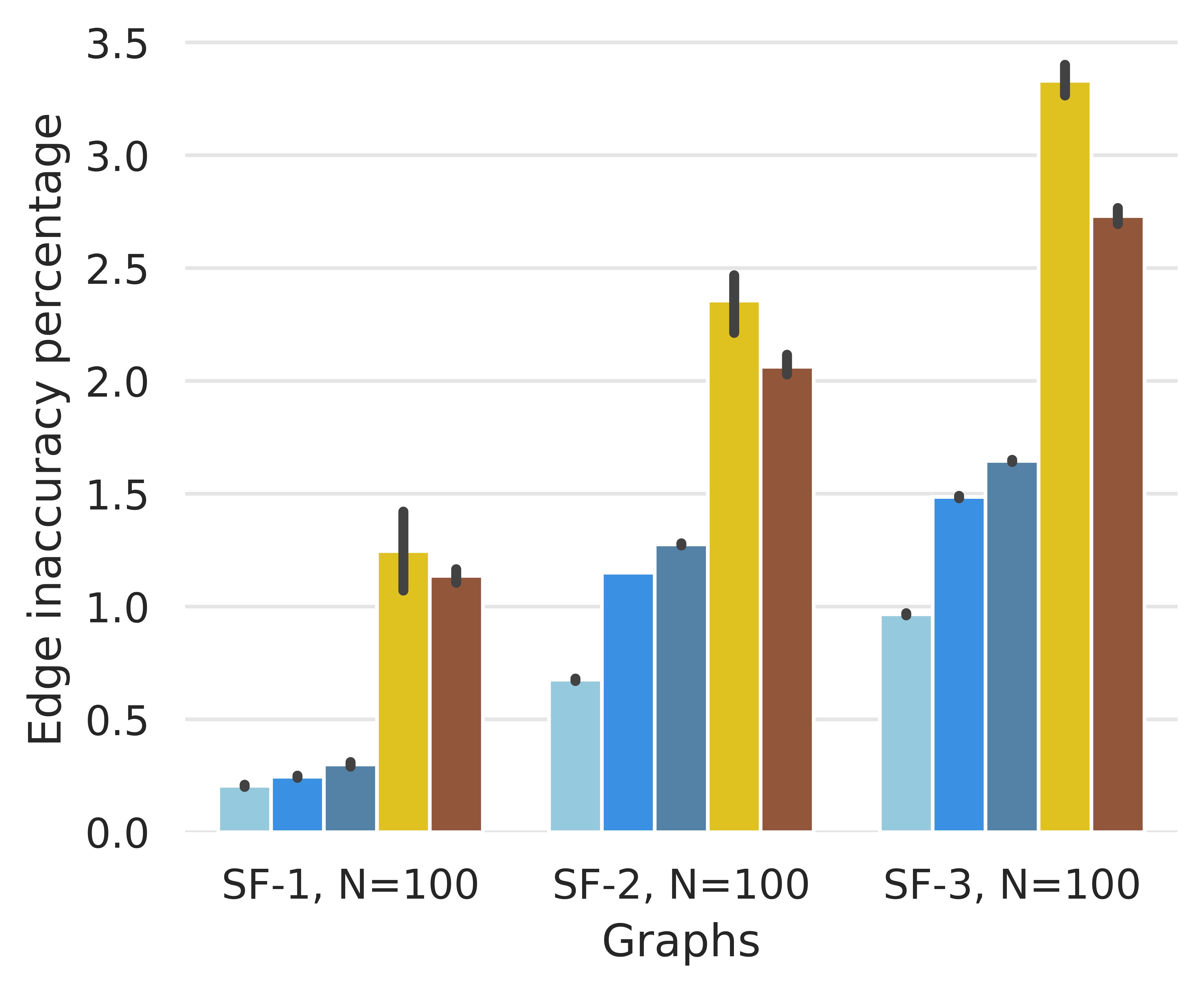}
        \caption{}
    \end{subfigure}
\hfill%
\begin{subfigure}[t]{0.18\textwidth}
\vskip 0pt
        \includegraphics[width=\textwidth]{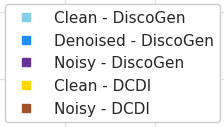}
    \end{subfigure}%

\caption{Results on clean (technical noise free), denoised, and noisy (with technical noise) data. \textbf{(a)}: Area under precision recall curve (AUPRC) of DiscoGen compared to DCDI. Higher is better.  DiscoGen significantly outperforms DCDI on all data. \textbf{(b)}: Percentage of edge inaccuracies  of DiscoGen compared to DCDI. Lower is better. DiscoGen significantly outperforms DCDI on all data. }
\label{fig:N100_results}
\end{figure*}

Since GRNs tend to be sparse, we focus primarily on reporting AUPRC results, as they provide the most informative insights (detailed results on the other metrics are given in Appendix \ref{sec:app:results}). Figure \ref{fig:N100_results} presents results on AUPRC on clean (technical-noise free), noisy (with technical-noise), and denoised data. On clean data, DiscoGen achieves an AUPRC of over $0.8$ on GRN graphs with all sparsity levels, while DCDI achieves around 0.15 AUPRC. Thus, DiscoGen outperformes DCDI by around $430\%$. On noisy data, DiscoGen achieves close to 0.8 AUPRC on SF-1 data, and over 0.55 AUPRC on SF-2 and SF-3 data. In contrast, DCDI achieves less than 0.02 AUPRC, regardless of graph sparsity. This indicates that DiscoGen performs over $2000\%$ better than DCDI.

On denoised data (shown in Figure \ref{fig:N100_results}
, DiscoGen achieves AUPRC over 0.8 on SF-1 data, and over 0.6 on SF-2 and SF-3 data. Notably, DiscoGen achieves better results on denoised data compared to noisy data, which is a strong indication that the denoise and compress method is effective in retaining useful information for inferring the GRN graph. 

\subsection{Ablation Results}

\begin{figure*}[t!]
\centering
\begin{subfigure}[t]{0.46\textwidth}
       \includegraphics[width=\textwidth]{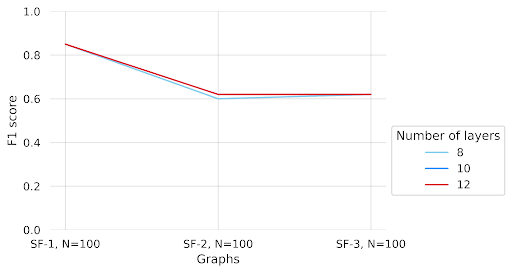}
        \caption{AUPRC for DiscoGen with 8, 10 and 12 layers of attention for \CSIVA. There is no significant difference between using 8 to 12 layers.}
    \end{subfigure}%
\hfill%
\begin{subfigure}[t]{0.46\textwidth}
        \includegraphics[width=\textwidth]{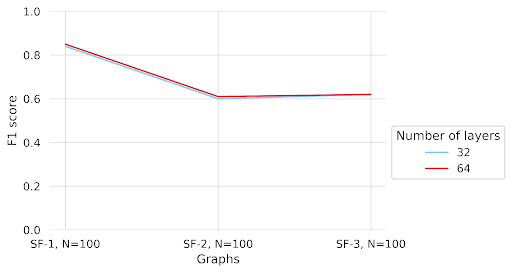}
        \caption{AUPRC for DiscoGen with $32$ and $64$ input embedding size for \CSIVA. There is no significant difference between using $32$ or $64$ hidden states.}
    \end{subfigure}

\caption{\textbf{AURPC for DiscoGen with different number of layers and of input embedding size.} 
}
\label{fig:ablation}
\end{figure*}

Our next investigation sought to determine sensitivity of DiscoGen's performance to hyperparameters choice. One important consideration is the impact of allocating more or fewer resources to \CSIVA. By default, \CSIVA~is trained using an encoder transformer with 12 layers and an input embedding size of 64. We first examined how performance is affected by reducing the number of layers in the encoder transformer. We trained \CSIVA~on 8, 10, and 12 layers and report the results in Figure \ref{fig:ablation} (Left)
. Surprisingly, we found that reducing the number of layers from 12 to 8 has little impact on performance. Next, we looked at how performance is affected by reducing the size of the embedding layers from 64 to 32. The results are shown in Figure \ref{fig:ablation} (Right). We found that even halving the embedding size has only a minor impact. These experiments suggest that DiscoGen is robust to hyperparameters changes such as the number of layers and hidden dimensions. Therefore, users could potentially use fewer layers and hidden dimensions if computational resources are limited.

Finally, we investigated how changing different hyperparameters of the data impacts DiscoGen's performance.  Specifically, we addressed the following questions: 1) What is the impact of  changing the number of replicates $K_n$? 2) What is the impact of changing the number of genes $N$?

\begin{figure*}[t!]
\centering
\begin{subfigure}[t]{0.32\textwidth}
       \includegraphics[width=\textwidth]{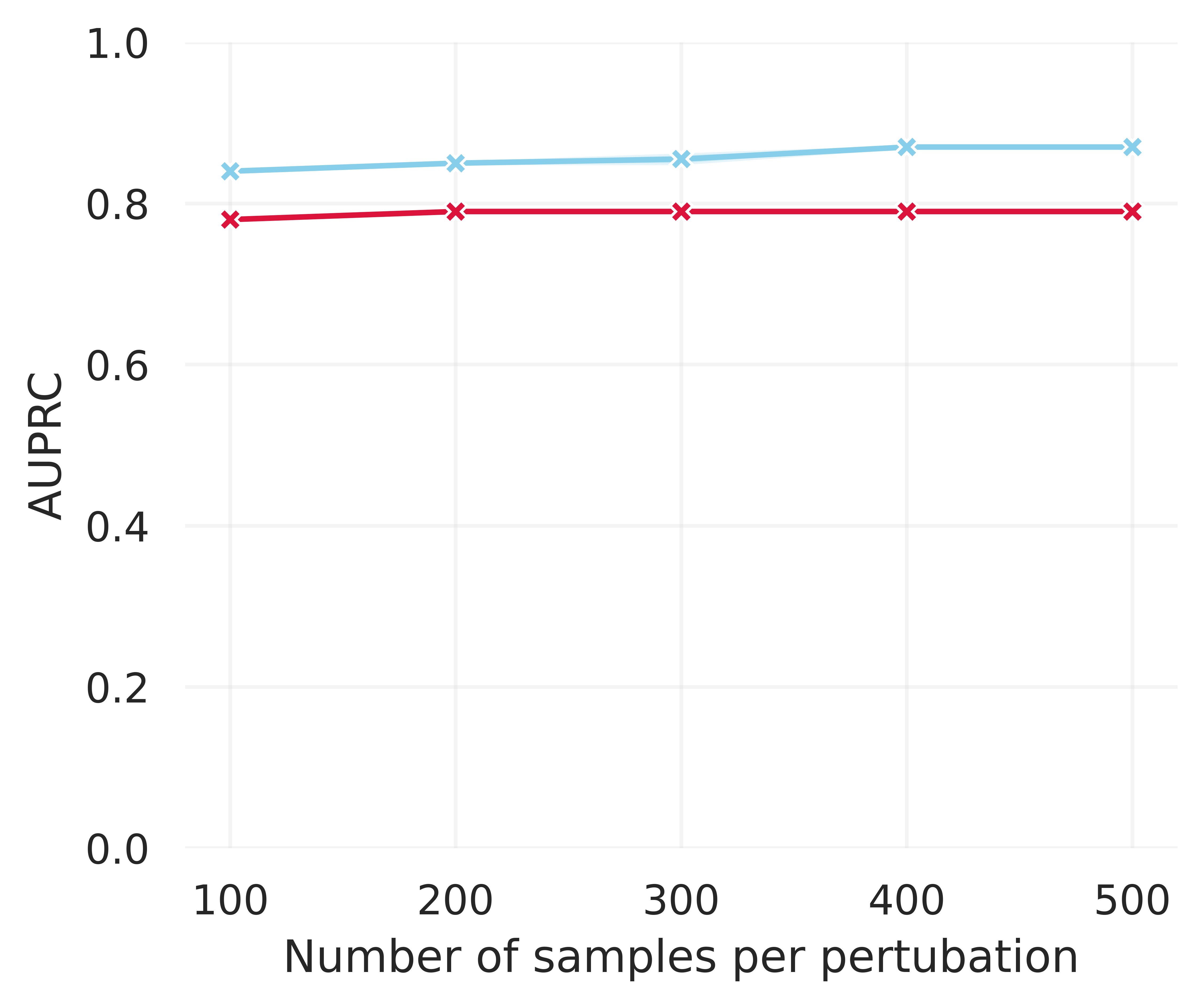}
        \caption{Scale-free graphs of degree 1 (SF-1).}
    \end{subfigure}%
\hfill%
\begin{subfigure}[t]{0.32\textwidth}
        \includegraphics[width=\textwidth]{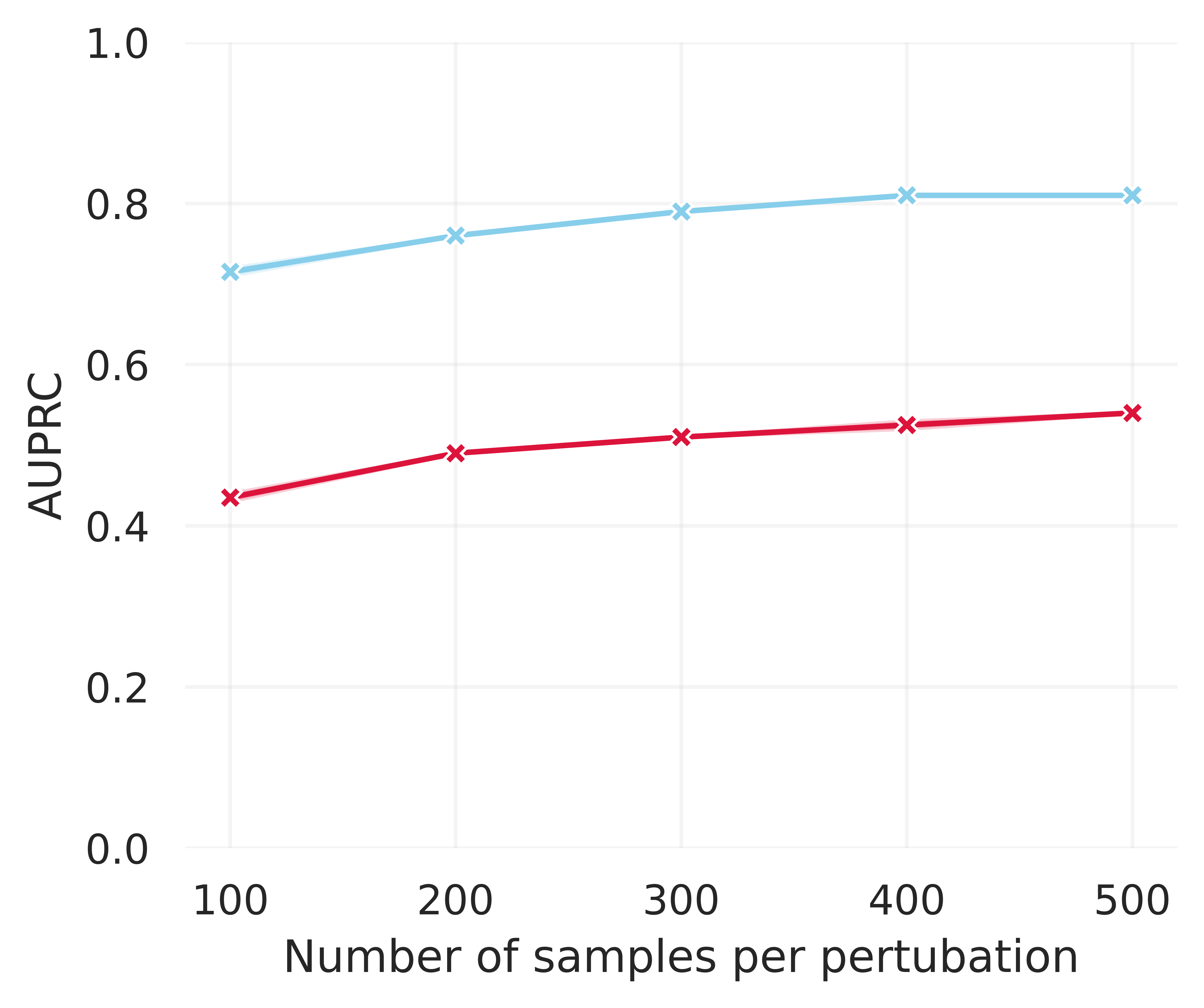}
        \caption{Scale-free graphs of degree 2 (SF-2).}
    \end{subfigure}
\hfill%
\begin{subfigure}[t]{0.32\textwidth}
        \includegraphics[width=\textwidth]{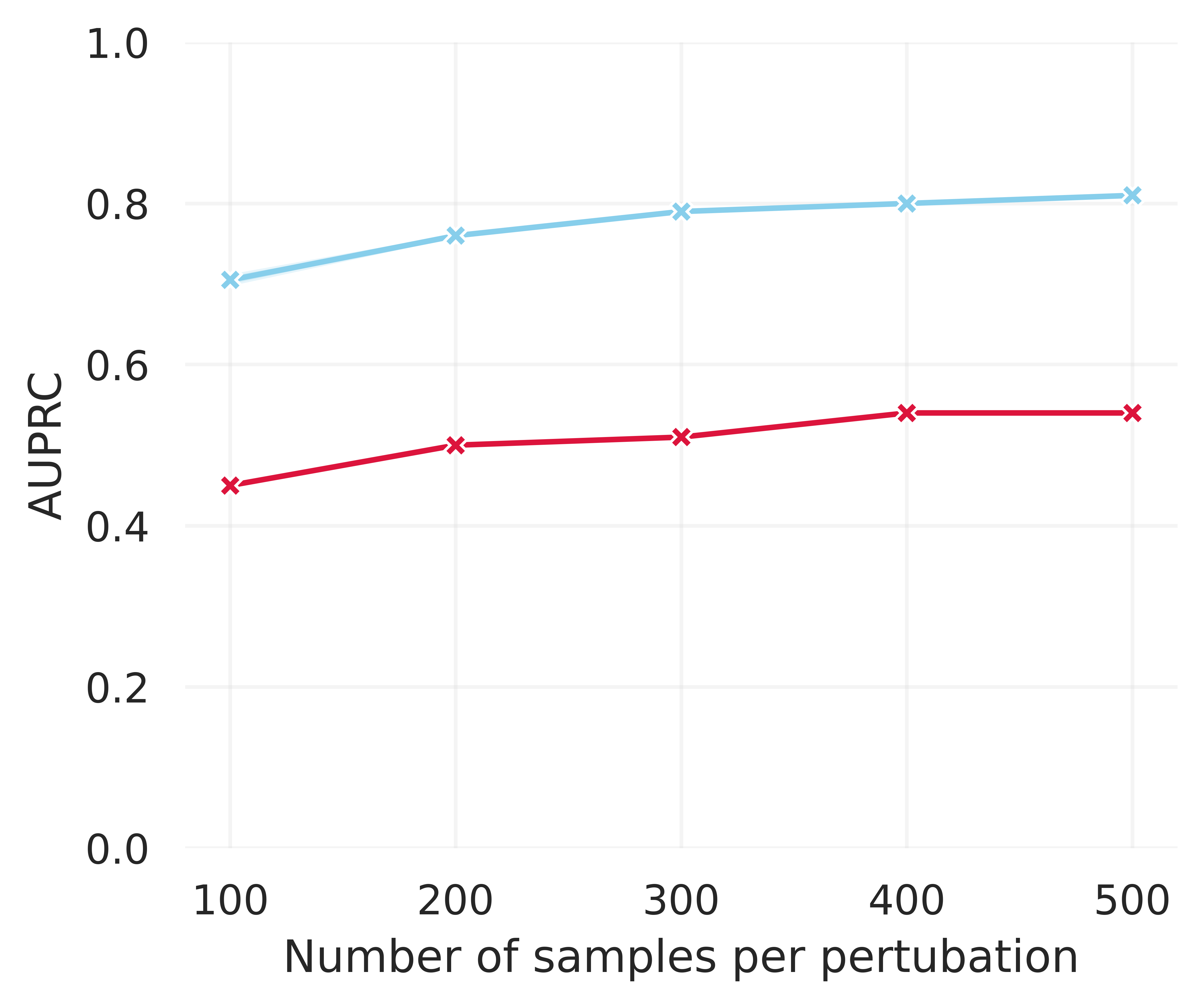}
        \caption{Scale-free graphs of degree 3 (SF-3).}
    \end{subfigure}%

\caption{AURPC for DiscoGen on number of replicates  per perturbation from $K_n=100$ to $K_n=500$, for clean (blue) (technical noise free) and noisy (red) (with technical noise) data. For denser graphs (SF-2 and SF-3), performance improves upon seeing more samples. }
\label{fig:different_num_samples}
\end{figure*}

To answer the first question, we varied the number of replicates per perturbation from the default setting of $500$ to $100$, $200$, $300$, and $400$. The results, shown in Figure \ref{fig:different_num_samples}, indicate that reducing the number of replicates from 500 to 300 only slightly decreases performance, while a more significant reduction is observed with only 100 replicates. These findings suggest that having more replicates generally improves performance, although the benefits are marginal beyond 300 replicates.

\begin{figure*}[t!]
\centering
\begin{subfigure}[t]{0.32\textwidth}
       \includegraphics[width=\textwidth]{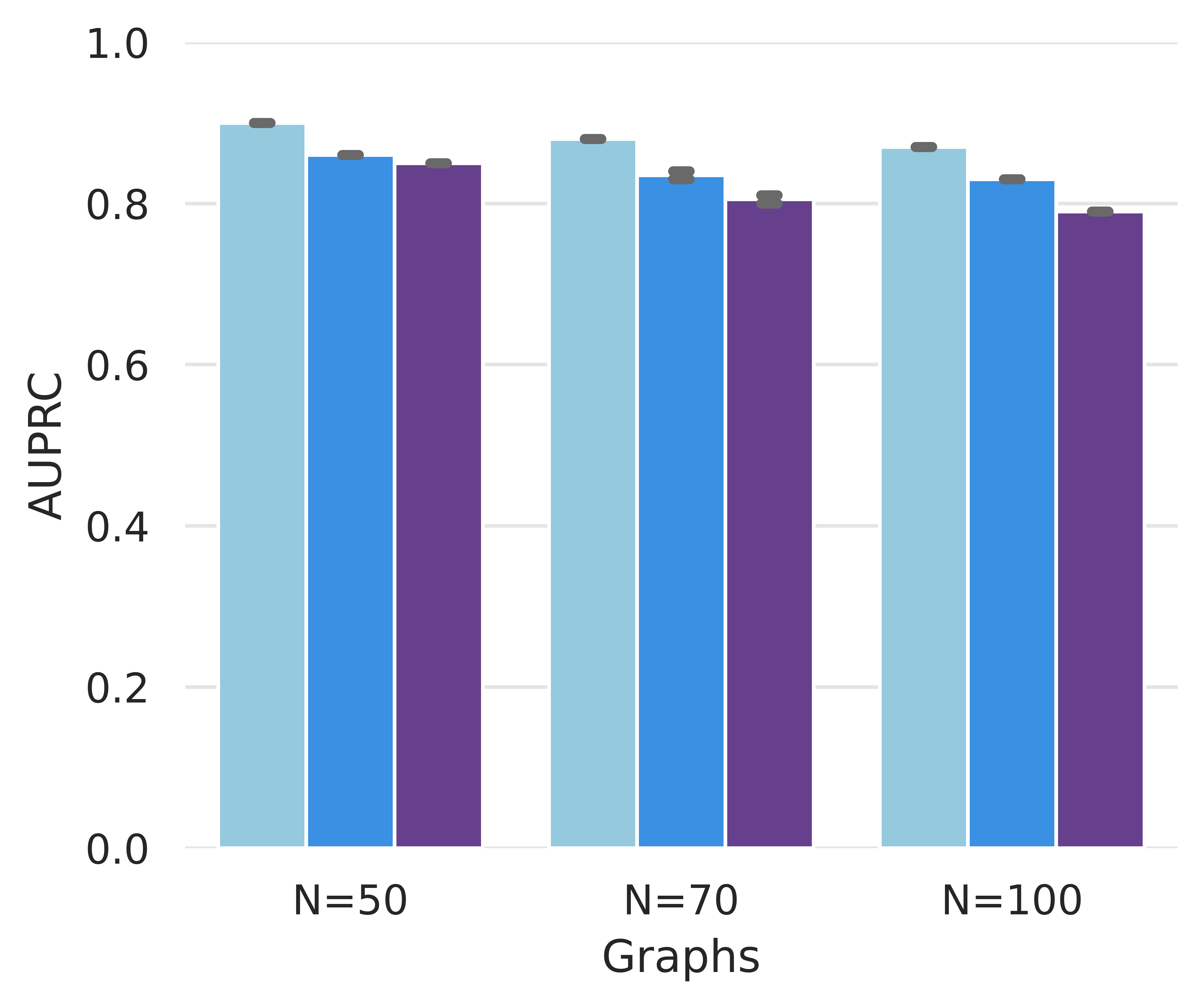}
        \caption{Scale-free graphs of degree 1 (SF-1).}
    \end{subfigure}%
\hfill%
\begin{subfigure}[t]{0.32\textwidth}
        \includegraphics[width=\textwidth]{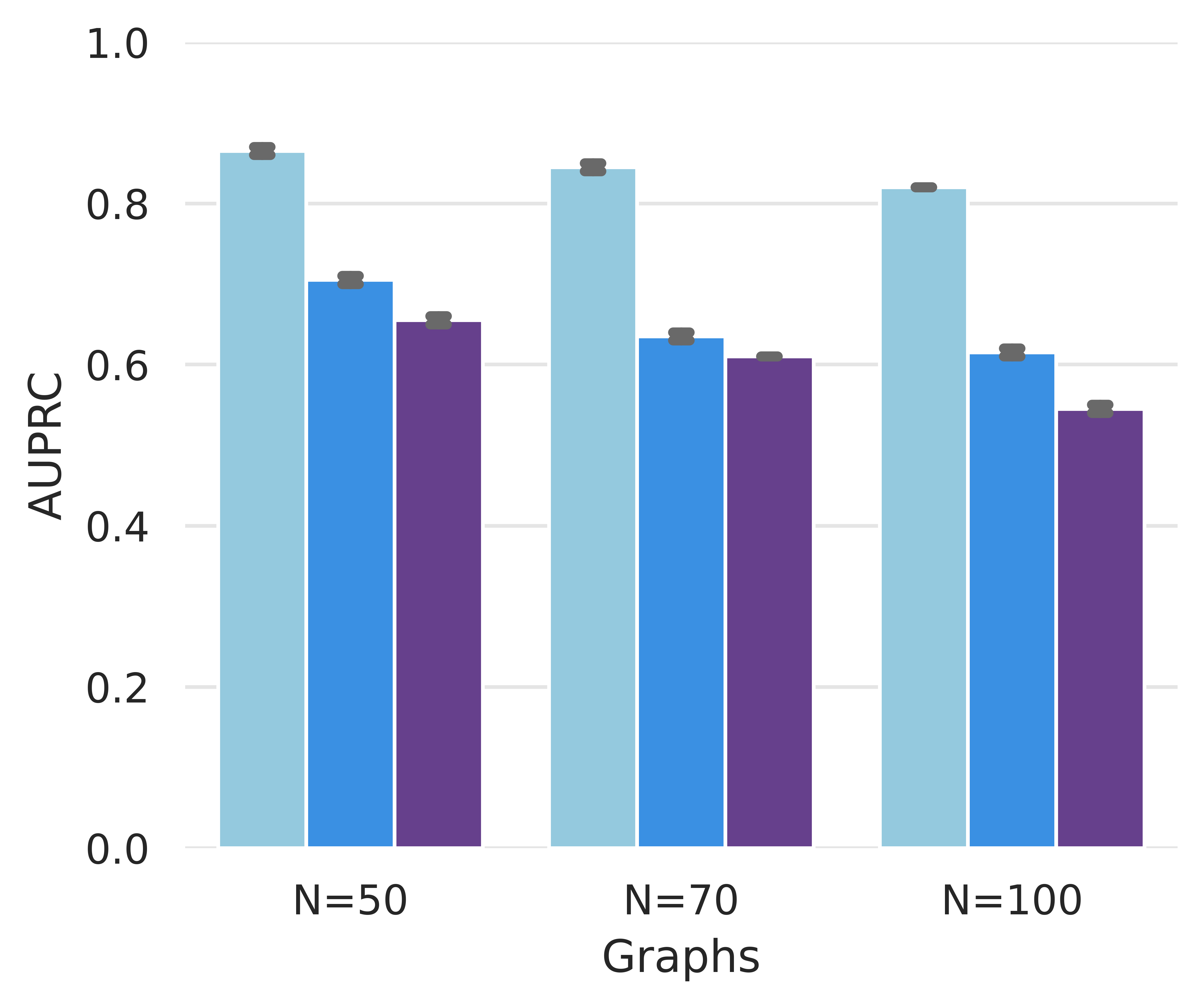}
        \caption{Scale-free graphs of degree 2 (SF-2).}
    \end{subfigure}
\hfill%
\begin{subfigure}[t]{0.32\textwidth}
        \includegraphics[width=\textwidth]{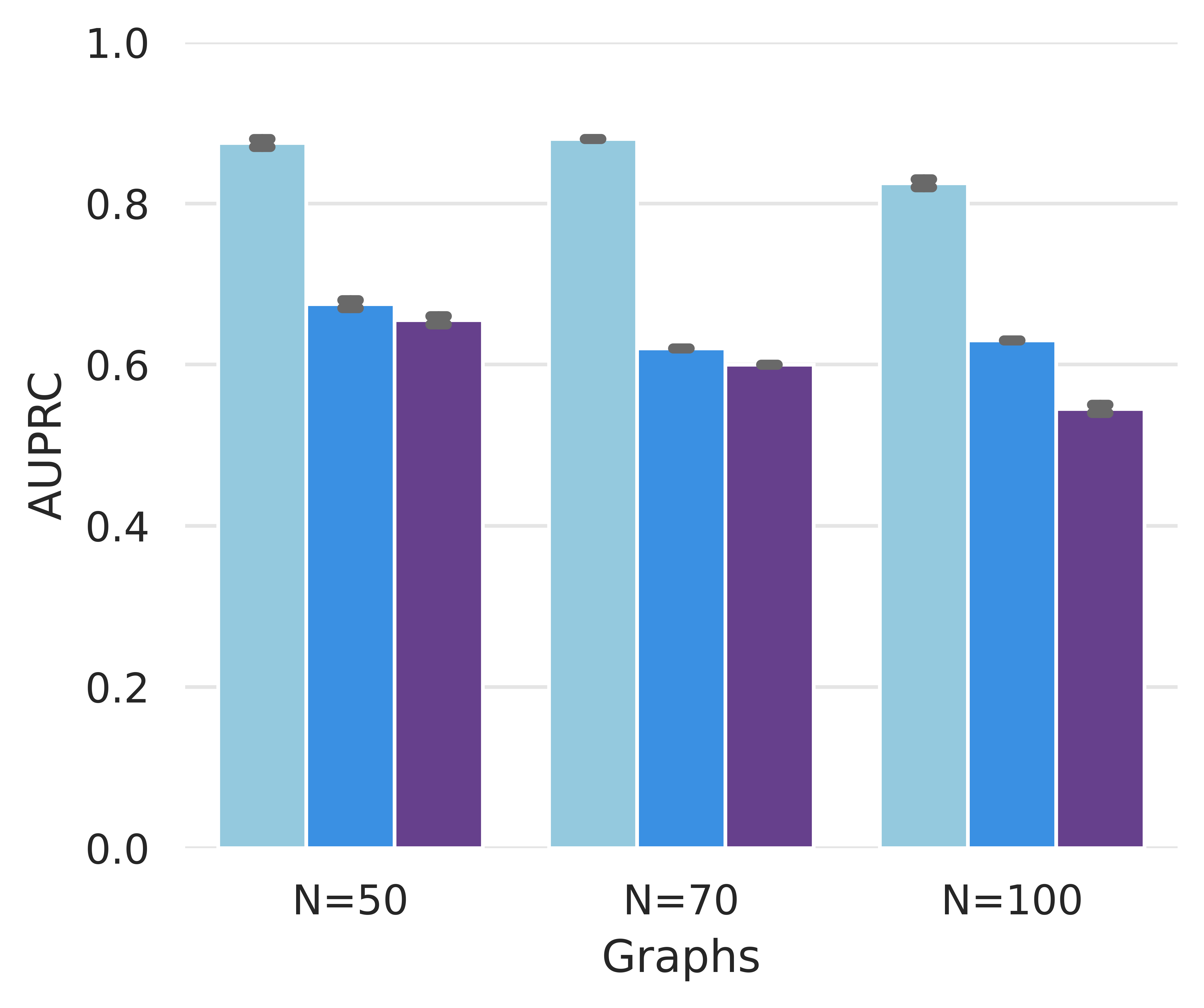}
        \caption{Scale-free graphs of degree 3 (SF-3).}
    \end{subfigure}%

\caption{\textbf{AURPC for DiscoGen evaluated on different number of genes, varying from  $N=50$ to $N=100$.} Models evaluated on clean (technical noise free) data, denoised and noisy (with technical noise) data. }
\label{fig:different_size_GRN_AUPRC}
\end{figure*}

To answer the second question, we varied the number of genes from $50$ to $100$. In causal discovery, larger graphs are typically more challenging to learn, as the number of possible graphs grows super exponentially with the number of nodes \cite{heinze2018causal}. The results are shown in Figure \ref{fig:different_size_GRN_AUPRC}. We can see that DiscoGen's performance worsen only marginally as the number of genes increases. This gives a strong indication that DiscoGen is reasonably robust to different number of genes.

\section{Methods}\label{online_methods}

\paragraph{Model architecture, training, and hyperparameters, and data setup}
The denoise and compress model is composed of a single-layer LSTM module with hidden dimension $64$, which outputs its hidden state to a mixture density network (MDN). The MDN module is a feedforward neural network with two hidden layers of dimension $32$ and $16$, respectively. The MDN's final layer generates a vector of size 6, which represents the coefficients of a Mixture of Gaussians with 2 mixture components. 
While it is possible to use different numbers of mixture components, we found that using 2 components works well. 
The model was optimized using the Adam optimizer with a learning rate of $1e-3$, and trained for $50k$ iterations.

The CSIVA model includes an encoder transformer with $10$ layers. To reduce computational complexity, we used MLPs for decoding instead of the large decoder transformer used in \cite{ke2022learning}. This approach has been previously used in \cite{ke2022learning} when training on large graphs. The model was optimized using the Adam optimizer with a learning rate of $1e-4$, and trained for $500k$ iterations. All inputs were first embedded using a feedforward neural network with a single hidden layer of size $64$ before being fed into the transformers. The transformer's positional encoding encodes the index of the gene. It's worth noting that loading data for large graphs with $100$ perturbations and $500$ replicates per perturbation can be slow. To mitigate this issue, we took the average of the replicates for the same perturbation in both the clean and noisy datasets before feeding them into the causal discovery model.

For data generation, we used the range indicated in \cite{dibaeinia2020sergio} for initialization of the SDE parameters and noise parameters. We run Sergio with $dt=0.01$ for $t=50.0$ to insures that the SDE has reached steady state. 


\section{Discussion}\label{discussion}
The development of DiscoGen represents a significant advance in the field of Gene Regulatory Network inference. The ability to handle perturbational data and denoise gene expression measurements are critical features for accurately modeling activatory and inhibitory interactions between genes. The results of our experiments shows that DiscoGen outperforms existing state-of-the-art neural network-based GRN discovery methods. Although the experiments are conducted on synthetic data, the promising results indicate that DiscoGen has the potential to be a valuable tool in biological research, and further validation on real-world biological data is warranted.

\bibliography{sn-article}
\clearpage

\begin{appendices}

\section{Hyperparameters}\label{sec:app:Hyper}
All of our experiments use hyperparamters in Table \ref{tab:app:hyper_denoising} and Table \ref{tab:app:hyper_csvia}. Hyperparameter search spaces for baseline DCDI \citep{brouillard2020differentiable,lopez2022large} are found in Table \ref{tab:app:hyper_dcdi}.

\begin{table}[h]
    \centering
    \caption{Hyperparameters for the denoise and compress model.}
    \begin{tabular*}{.6\textwidth}{@{\extracolsep{\fill}}lc}
    \toprule
      \textbf{Hyperparameter} & \textbf{Value} \\ 
      \midrule
      LSTM Hidden state dimension & 64\\ 
      MDN Hidden state dimension & [32, 16] \\ 
      Optimizer & Adam\\ 
      Learning rate & \(10^{-3}\) \\
      Number of random seeds & 3 \\
      Training iterations & $50,000$ \\
      \bottomrule
    \end{tabular*}
    \label{tab:app:hyper_denoising}
\end{table}

\begin{table}[h]
    \centering
    \caption{Hyperparameters for \CSIVA~\cite{ke2022learning}.}
    \begin{tabular*}{.6\textwidth}{@{\extracolsep{\fill}}lc}
    \toprule
      \textbf{Hyperparameter} & \textbf{Value} \\ 
      \midrule
      Hidden state dimension & 64\\ 
      Encoder transformer layers & 12 \\ 
      Num. attention heads & 12 \\ 
      Optimizer & Adam\\ 
      Learning rate & \(10^{-4}\) \\
      Number of random seeds & 3 \\
      Number of cells per perturbation (k) & 500 \\
      $S$ (number of samples) & $50,000$ \\
      Num. training datasets $I$ & 
      $50,000$  \\
      Training iterations & $500,000$ \\
      
      \bottomrule
    \end{tabular*}
    \label{tab:app:hyper_csvia}
\end{table}

\begin{table}[h]
    \centering
    \caption{Hyperparameters search spaces for DCDI \citep{brouillard2020differentiable}.}
    \begin{tabular*}{.6\textwidth}{@{\extracolsep{\fill}}lc}
    \toprule
      \textbf{Hyperparameter} & \textbf{Value} \\ 
      \midrule
      Hidden state dimension & \{16, 32, 64\}\\
      Number of hidden layers & \{1, 2\}\\
      Optimizer & Adam\\ 
      Learning rate & \(\{5e^{-3}, 1e^{-3}, 5e^{-4}, 1e^{-4}\}\) \\
      Number of random seeds & 3 \\
      \bottomrule
    \end{tabular*}
    \label{tab:app:hyper_dcdi}
\end{table}

\section{Additional Results}\label{sec:app:results}

\begin{figure}
    \centering
    \includegraphics[width=0.75\textwidth]{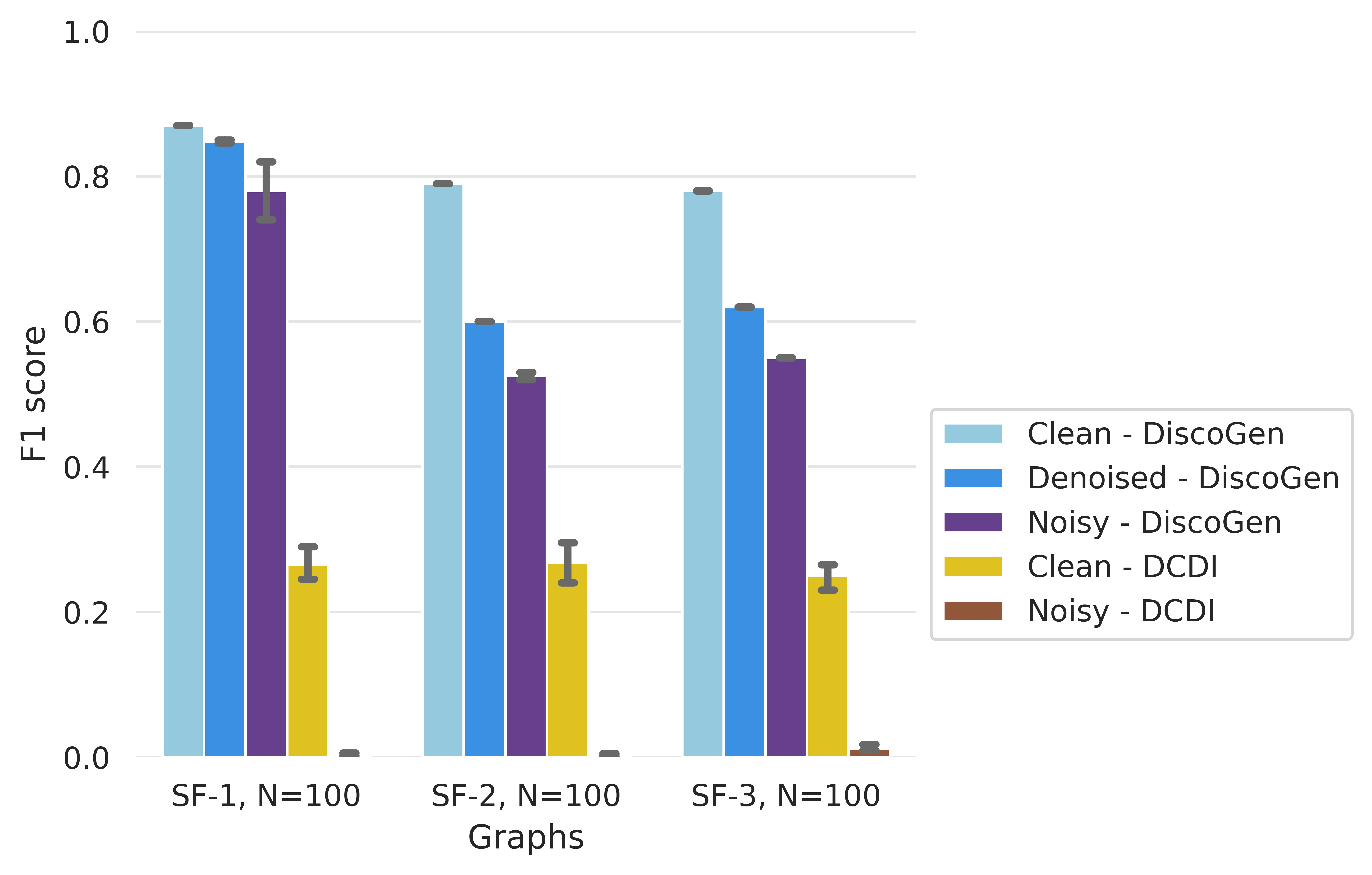} 
    \caption{F1 scores (higher is better) for DiscoGen compared to DCDI on SF-1, 2 and SF-3 graphs.}
    \label{fig:f1}
\end{figure}

We also evaluated models using F1 scores. The results are shown in Figure \ref{fig:f1}.  DiscoGen consistently outperforms DCDI on both clean and noisy data. Note that, DCDI's performance significantly deteriorates when applied to noisy data. This highlights the limitation of DCDI in handling noisy data, while DiscoGen is much more robust in such scenarios.

%




\end{appendices}

\end{document}